\documentclass[12pt,preprint]{aastex}

\slugcomment{To appear in the Astronomical Journal}

\shorttitle{The Morpholgy of IRC~+10420}
\shortauthors{Tiffany et al.}

\begin{document}


\title{The Morphology of IRC~+10420's Circumstellar Ejecta\footnote{Based on observations with 
the NASA/ESA {\it Hubble Space Telescope} obtained at the Space Telescope Science Institute, which 
is operated by the Association of Universities for Research in Astronomy, Inc. under NASA contract NAS 5-26555.}}


\author{Chelsea Tiffany, Roberta M. Humphreys, Terry J. Jones and Kris Davidson }
\affil{Astronomy Department, University of Minnesota, Minneapolis, MN 55455 }

\email{tiffany@astro.umn.edu, roberta@umn.edu}

\begin{abstract}
Images of the circumstellar ejecta associated with the post-red supergiant IRC~+10420 
show a complex ejecta with visual evidence for episodic mass loss. In this paper we describe
the transverse motions of numerous knots, arcs and condensations in the inner ejecta measured
from second epoch {\it HST/WFPC2} images. When combined with the radial motions for several of
the features, the total space motion and direction of the outflows show that they were ejected
at different times, in different directions, and presumably from separate
regions on the surface of the star. These  discrete structures in the ejecta are kinematically 
distinct from the general expansion of the nebula and their motions are dominated by their
transverse velocities.  They are  apparently all moving within a few degrees of the plane of the 
sky. We are thus viewing IRC~+10420 nearly {\it pole-on} and looking nearly directly down onto 
its equatorial plane. We also discuss the role of surface activity and magnetic fields on
IRC~+10420's recent mass loss history.
\end{abstract}


\section{Introduction}

IRC~+10420 is one of the most important stars in the upper HR diagram for understanding the final stages of massive star evolution.  With its high luminosity ($L \ {\sim} \ 5 {\times} 10^5$ $L_{\odot}$) and A-F--type spectrum it is one of the stars that defines the empirical upper luminosity boundary for evolved stars in the HR diagram \citep{HD94}. IRC~+10420 is also a strong OH maser, one of the warmest known, and  one of the brightest 10--20 $\mu $m IR sources in the sky with an With its extraordinary mass loss rate (3 -- 6 $\times$ 10$^{-4}$ $M_{\odot}$ yr$^{-1}$)\citep{Knapp85,Rene96,RMH97}. 

  It has been variously described in the literature as either a true supergiant 
  \citep{RMH73,Gig76,Mut79} 
  or  a proto-planetary/post-AGB star \citep{Hab89,Hri89,Bow89},
  depending on distance estimates that ranged from 1.5 to 7 kpc. 
  \citet{TJJ93} (Paper I) combined multi-wavelength spectroscopy, photometry, 
  and polarimetry to confirm a large distance of 4--6 kpc and the 
  resulting high luminosity mentioned above.  This  conclusion was  supported by 
  \citet{Rene96} who demonstrated from CO data that IRC+10420 has
  to be much more luminous than  the AGB limit.  
  
  {\it HST/WFPC2}  images  (Humphreys et al 1997, Paper II) revealed a complex circumstellar environment with numerous small condensations or knots, ray-like features, and intriguing semi--circular arcs or loops.  A few other intermediate-temperature hypergiants such as $\rho$ Cas and HR 8752 occupy the same region in the HR diagram, but IRC +10420 is the only one with apparent circumstellar nebulosity \citep{MTS06}, making it our best candidate for a star in transition from a red supergiant to an S Dor--type variable (LBV), a Wolf-Rayet star, or a pre--supernova  state.  IRC~+10420 has shown some significant changes during the past century. It brightened by a magnitude or more in the 50 years prior to 1970 (Gottleib \& Liller 1978, Paper I) and its apparent spectral type changed from a late F to an A-type supergiant in just the past 30 years \citep{Rene96,Rene98}, although  Humphreys et al (2002), hereafter Paper III, demonstrated that  IRC~+10420's  wind is optically thick.  Consequently, the observed variations in apparent spectral type and inferred temperature are more likely due to changes in its wind,  and not to interior evolution on such a short timescale.

The HST images of IRC~+10420 show no obvious axis of symmetry, although the complex ejecta  provide evidence for more than one high mass loss episode. The outermost reflection arcs at $\approx$ 5$\arcsec$ were ejected about 3000 years ago, possibly when the star was still a red supergiant, while the  very complex structures closer to the star correspond to much more recent asymmetric mass loss events. Surface photometry of the optical and near-infrared images (Paper II) showed that IRC +10420 experienced  a high mass loss episode during the past 600 years, shedding about 1 $M_{\odot}$ with a mass loss rate $\sim$ 10$^{-4}$ $M_{\odot}$ yr$^{-1}$.  \citet{Block99} suggested that a high mass loss episode ended $\sim$ 60--90 years ago which interestingly corresponds to the apparent onset of its brightening.  The numerous arcs, knots, and jetlike structures are suggestive of localized ejection events in seemingly random directions which may be due to large scale active regions on the star (Paper III).

HST/STIS long-slit spectroscopy of the reflection nebula allowed us to effectively view the star from different directions (Paper III).  The extracted spectrum at each location is essentially that of a reflection or scattering nebula.  Measurements of the absorption minimum in the strong double-peaked H$\alpha$ emission profile showed that the reflection nebula is expanding in a spherical outflow at $\sim$ 60 km s$^{-1}$.  Where a slit crossed one of the semi-circular arcs the velocities deviate from the expansion of the surrounding nebulosity indicating that the arc is a kinematically separate feature.  The shape of the H$\alpha$ emission profile was remarkably uniform throughout the ejecta, contrary to what we would expect from previous models with an equatorial disk (Paper I) or bipolar outflow (Oudmaijer et al 1994, Paper II). More recent observations from integral -field spectroscopy 
\citep{Davies07} provide  additional evidence for a bipolar outflow, although the signature is not strong
close to the star. Recent near-infrared interferometry \citep{deWit08} reveals an elongated emitting region on the milli-arcsecond scale about twice the size of the star. However their results are not conclusive as to 
whether it is an edge-on disk or bipolar outflow. One possible explanation for the apparent conflicting evidence for the ouflow and structure of IRC~+10420's circumstellar ejecta may be due its orientation with respect to our line of sight. We could be viewing IRC~+10420 nearly pole-on.

We have now obtained second epoch Planetary Camera images of IRC~+10420 to measure the transverse motions of the discrete structures in the ejecta. In combination with radial velocities from the previous long-slit spectroscopy, we can determine when the material was ejected, a history of IRC~+10420's episodic mass loss, and the morphology of its ejecta based on direct observations. The orientation and vector motions of these features will be important for understanding the origin of the ejecta and the responsible mass loss mechanism.  In the next section we describe the observations, data processing and measurement procedure for the transverse motions. In \S {3} we discuss the motions of the 
separate arcs and knots. We  summarize  the results on the geometry of the ejecta and 
IRC~+10420's mass loss history in \S {4} and \S {5} and conclude with remarks on the role of
convective activity and magnetic fields in \S {6}. 

\section{The HST Observations, Data Processing, and Measurement Procedure}

\subsection{Observations}

%Data information
HST/WFPC2 images of IRC~+10420 were obtained on April 5, 1996 (Epoch 1) and on March 9, 2008 (Epoch 2)
with the Planetary Camera  giving a baseline of 11.93 years. The F467M and F547M filters used in Epoch1  were repeated in Epoch 2 plus additional images with the F675W and F1042M filters.  Because of the wide dynamic range of the nebulosity associated with IRC~+10420, a range of exposure times was necessary to sample the complex ejecta. For ease of comparison, similar exposure times were used for 467M and F547M. The complete list of filters, exposure times and the subsequently combined images are given in Table 1. A number of images were taken using the Dither mode of the HST. This allows images with matching filters and exposure times to be taken with a 2.5 or 5 pixel offset, which helps identify bad pixels on the chip as well as giving better image detail in the final combined images. The dithered images are noted on Table 1.

%Image reduction
Both sets of data were reduced in tandem, using IRAF packages, to assure consistency. We used routines  in the HST Dither Handbook \citep{Dither}  for a diffuse source imaged with the PC2. Cosmic rays were removed with the task {\it precor},  and  the shift between images in the same  epoch were measured with \texttt{crossdriz} and \texttt{shiftfind}. Image masks were created from the bad pixel information from HST and from the images themselves. The final images were created using these masks and the shifts to drizzle the images together.  The images were then subsampled by a factor of two for a  final resolution of .02275$\arcsec$/pix. Images from the same epoch were then aligned and combined.  With this resolution  at a distance of 5 kpc and 
11.9 year baseline, we expect to be able to measure motions as small as $\sim$ 15 km s$^{-1}$. 

%PSF removal and final alignment
The next step was to remove the Point Spread Function (PSF). We followed the 
procedure described in Paper II. We used TinyTIm to create  simulated  HST specific PSFs 
to match the WFPC2 observations. However, TinyTIM has several limitations that affect its
application to IRC~+10420. To produce an accurate PSF for the PC chip, the PSF was sub-sampled by two, but this decreased the size of the PSF which was already too small for IRC~+10420. For IRC~+10420 an input size of 18$\arcsec$ is needed.  Furthermore, the TinyTIM PSF is
color dependent with a maximum allowed $B-V$ color of 1.6 mag. IRC~+10420 is highly reddened by interstellar extinction ($B-V$ of 2.7 mag); consequently we adjusted the color table 
to  match the spectral energy distribution  of IRC~+10420 as closely as possible. The TinyTIM PSFs were then smoothed  to match the images. The IRAF task \texttt{lucy} was used to deconvolve the images. Even with the reddened PSF, matching and removing the diffraction spikes for IRC+10420 was only partially successful. When the bright central area was fully subtracted, the diffraction spikes in the extended nebulosity were oversubtracted, creating artifacts. These were easy to recognize because the artifacts were aligned with the diffraction spikes, which were not aligned between the two epochs. To avoid creating artifacts we reduced the number of iterations to ten.  To more 
accurately remove the diffraction spikes in the regions of the knots and arcs we would be measuring, we  undersubtracted the innermost  region. In the 547M long exposure combinations (547l1 and 547l2), this central area was completely saturated. It was replaced with the center of the corresponding short image (547s1 and 547s2 respectively) and scaled to match the long exposure time to simulate the correct profile for the deconvolution routine to remove the diffraction spikes.  A precision alignment was then done with the aid of field stars on the edges of the images.

\subsection{Measurement Procedure}
%General Procedure
We combined the  multiple F467M and F547M exposures at each epoch into a short ({\it s}) and long ({\it l}) exposure for measurement. The combined short and long images in each filter pair were then blinked using DS9 to identify features common to both epochs. Not all of the features were equally visible in both filters and in both epochs. The images could not simply be cross-correlated due to the non-uniform 
background and the residuals from the PSF subtraction. Each individual knot, condensation or arc would have required its own cross-correlation which would have been difficult with the variable background. We therefore blinked the aligned images to find observable motion. The center point and position angle of each feature was then measured relative to the central star. This procedure was repeated for each filter combination in which the feature was present. With two filters, short and long, there  were four possible combinations. We then repeated these measurements three times, cycling through the four filter combinations in a different order with each set separated by several days to avoid measurement bias.  Because of the complications described above with the PSF subtraction we avoided measuring any features near the residual diffraction spikes and within a radius of $\sim 0\farcs33$ from the star to avoid confusion with the deconvolution of the star. The diffraction spikes were rotated by $\sim$ 16 degrees between the two epochs, so we can be confident that none of the features we measured are associated with them. 

The positional offsets in x and y between the two epochs were then determined for each feature. The angular offsets  with the number of measurements are in Table A1 in the Appendix. In the following analysis and discussion we use  our 
preferred distance of 5 kpc (see Paper I). In Table 2 we give the 
projected radial distance and the position angle relative to the star in the plane of the sky for each 
measured feature with its corresponding mean transverse velocity, at the adopted  distance of 5 kpc, and its direction of motion ($\phi$) with their mean errors. Figure 1 shows a short exposure image (547s1) of IRC+10420  with the outlines of the general regions discussed in this paper. The complex inner two arcsec region is shown in Figures 2a and 2b 
with individual knots and condensations identified.  Figure 3 shows an 
enlargement of a section of one of the semi-circular arcs to illustrate the offset 
between epoch 1 and epoch 2. 

%Cross-correlation with outer feature
The Outer SE Feature (see Figure 1) was the one
exception to the measurement procedure described above. This large, diffuse feature has no well defined bright center or condensation, so we used cross-correlation to determine the offset.  This worked for this filamentary feature because it is far from the central star and the diffraction spikes, with no significant background. It also has a distinct 'V' shape that is easily identifiable between epochs and can
be matched using cross-correlation. This was done twice with different sized regions ($\sim$ 1.5$\arcsec$ and $\sim$ 2.5$\arcsec$) around the feature to verify the cross-correlation. No other features in the outer regions are prominent enough to be measured this way.

\section{The Kinematics of the Ejecta}

Radial velocities, measured  from the absorption minimum in the very strong  H$\alpha$ line are available for numerous positions along two separate slits from the long-slit spectra obtained with HST/STIS in 1999 (Paper III). The slits cross several of the brighter knots and arcs near the star, including one of the semi-circular arcs.  The observed spectrum is that of an expanding reflection nebula and the apparent 
 Doppler velocity at each reflective condensation is due to the velocity of the star when it is observed directly, the expansion velocity of the nebula, and the relative velocity component along our line of 
 sight of the condensation. Figure 9 in Paper III shows the measured Doppler velocity increasing with increasing distance from the star. The relation between the measured velocities and the three dimensional position can be fit by an expansion of 50 km s$^{-1}$ (Paper III), consistent with the outflow velocities from the CO and OH observations and the double-peaked hydrogen and Ca II emission. Adopting this model for the expansion of the nebula we  can estimate the relative motions of  the knots and arcs along the line of sight.  The relative radial component is  combined with the measured transverse velocity to determine the total space motion and the combined direction of motion  relative to the plane of the sky for individual knots. The results  are included in Table 3. In the subsequent discussions, all of the expansion ages or time since ejection are determined assuming a constant speed since ejection. For those objects with only a transverse motion, the expansion age is an upper limit. The resulting ages are identified and listed separately to differentiate them from the ages found from the total velocity.

The results for the individual features identified in Figures 1 and 2  are 
summarized in Tables 2 and 3. In the 
following subsections we discuss a few of the major features, the East jet, the SW
fan, the SW square,  and the semi-circular arcs.  

\subsection{The East Jet }

The East Jet (Figure 2b) is one of the more visible features in the ejecta. It  appears to be a row of several knots aligned almost directly east of the star.  In the first epoch image the knots are quite close to a diffraction spike, so the innermost knots closest to the star were not measurable  between the two epochs. Knots B and C have  transverse velocities of  165 km s$^{-1}$  and  136 km s$^{-1}$. Their respective times since ejection are 100 and 150 years, indicating that they are from a common recent ejection event, similar to SE Jet(2) Knot A. Knot C also has a small radial motion
(-5.9 km s$^{-1}$).  It is thus moving at 136 km s$^{-1}$ essentially in the plane of the sky at an angle of only 2.5$^{\circ}$ towards us.

\subsection{The South West Fan}
The scalloped or rippled pattern to the SW of the star that we call the fan is very prominent in the images of IRC~+10420 (see Figure 1).  However, the complexity of this region makes it hard to compare the diffuse condensations between the two epochs with a high degree of confidence.  Knots A and B (Figure 2a) are sufficiently distinct and bright enough that they can be identified in the images from both epochs.  Knots A and B have essentially the same  transverse velocities ($\approx$ 100 km s$^{-1}$) and a corresponding expansion age of 450 -- 470 years.  Both knots are moving nearly radially away from the star.  Knot B also has a  radial motion
of -15.3  km s$^{-1}$ yielding a total velocity of 104 km s$^{-1}$  and an ejection time of 450 yrs ago.  The fact that these knots have such similar properties, despite being separated by 1.5 arcsec, suggests that the fan is a physically distinct feature from a single mass loss event

\subsection{The South West Square and The South West Jet}
The South West Square is a grouping of 4 knots 1$\arcsec$ - 1.3$\arcsec$ from the star (Figure 2a) with position angles from  -125$^{\circ}$ to -137$^{\circ}$.  Knots B, C and D have similar velocities (89 - 107 km s$^{-1}$) as well as similar directions of motion (from -160$^{\circ}$ to -174$^{\circ}$) which are more or less radially away from the star. Their expansion ages from their transverse motions range from 250 - 360 years. Knot C also has a radial velocity of -19.5 km s$^{-1}$ which gives a total velocity of 98 km s$^{-1}$ and an expansion age of 320 years. These results suggest a common ejection event for these 3 knots.  

The South West Jet only has one measurable knot.  It is moving radially outward at 218 km s$^{-1}$ with an expansion age of only 70 years.  Its small associated radial motion of -5.6 km s$^{-1}$ does not alter its space motion or time since ejection.

\subsection{The Semi-Circular Arcs}
Three very intriguing nearly circular arcs can be easily seen to the east of the star (Figure 1). The arcs appear to be made up of several small knots, identified in Figure 4, that form the larger arcs or loops. Given their appearance, it is possible that these knots may have been part of three initially much more compact features, and the nearly circular arcs we observe now are actually expanding bubbles or loops. 

We initially measured the motions of the individual knots with respect to the star as we did for the other condensations discussed in this section (Table 2). The six knots in Arc 1 have a mean transverse velocity of 97 km s$^{-1}$ and a corresponding mean time since ejection of $\approx$ 400 years. Given its location, Knot A may be a separate condensation simply projected onto Arc 1. Excluding it gives 103 km s$^{-1}$ and 370 years. Radial velocities are available for three  of the knots (Table 3) giving a mean total velocity of 111 km s$^{-1}$ and an expansion age of 320 years. We note that the three knots  are consistent with a mean orientation 
that places them essentially in the plane of the sky, although there may be a slight tilt to the arc. Knot A in Arc 2 appears to be on the inside edge of the arc, is difficult to measure and gives an indefinite result.  Excluding Knot A, the other five knots give a mean transverse velocity of 91 km s$^{-1}$ and a corresponding mean time since ejection of 500 years. Knot F has an associated radial velocity and a corresponding expansion age of 400 years. Arc 3 is the least well-defined in the images, and  we were able to measure only three knots. Knots B and C in Arc 3 give consistent results with a mean transverse velocity of 150 km s$^{-1}$ and an expansion age of $\approx$ 300 years. 

The arcs are all about the same radial distance from the star, and the results for the individual knots suggest that they may all have been ejected at about the same time, 300 -- 400 years ago.

% How we did the measurement of the center point
To investigate whether the arcs are expanding bubbles, we examined the motions of the knots with respect to the center of their respective arcs. We first found the center of each arc by fitting an ellipse to the positions of the knots in DS9. We experimented with different ellipticities and adopted the
best fit to the measured postions of the knots that also best mapped the inter-knot diffuse nebulosity. This was done independently for each epoch. The percentage difference between the positions of the knots and the elliptical fit was used to determine the quality of the  fit. The center of the best-fit ellipse was then adopted as center of each arc. The corresponding transverse motion between the two epochs is summarized in Table 4{\footnote{In Arc 2, Knot A was not used to determine the best-fit position and was not included in the percentage difference calculations.}}. While we were able to determine a best fit, the adopted error for the velocity of each arc center is from the range in the transverse velocity derived from the different fits. 

% How we found the radial velocity for the center point of Arc 1.
Radial motions are available for three knots in Arc 1. Consequently we were able to solve for the velocity of the arc center assuming that the velocities of the knots, relative to the star, depend on the radial velocity of the center and their x and y positions relative to the arc center. The resulting radial velocity for the center of Arc 1 is essentially zero km s$^{-1}$, from both Epochs 1 and 2,  giving  a total space motion of 51 km s$^{-1}$ and a time since ejection of 700 yrs. 

%Arc 1
With this information for Arc 1 we then determined the tranverse motion of the individual knots relative to the arc center and for knots B, C, and G, with radial velocities, their total motions and orientation or tilt relative to the arc center. These results are summarized in Tables 5
and 6. The results confirm that Arc 1 is slightly tilted with respect to our line of sight. We also find that the expansion age or time for the arc of $\sim$ 100 yrs is significantly less than the time since ejection from the star whether we use the $\approx$ 700 yrs for the arc center or the $\sim$ 320 yrs determined from the three arcs with total motions (Table 3).  

%Arc 2
The tranverse motion for the  Arc 2 center gives a time since ejection of 370$^{+200}_{-100}$ years. Only Knot F in Arc 2 has a measured radial velocity; consequently, there is no radial velocity estimate for the arc center. The transverse motions for the individual knots give an average expansion time for the arc of $\sim$ 200 years. Although  there is a large spread in the results for the knots (Table 5), the results for the expansion of the arc itself and the ejection times for the knots  are consistent within the measurement uncertainties. 

%Arc 3
Arc 3 has an expansion age with respect to  the star of 300$^{+200}_{-100}$ years from its 
transverse motion. No radial velocity information is available. The expansion time of 50 years 
for the knots relative to the arc center (Table 5) is significantly less than the time since ejection from the star, although motions  are available for only three knots.  Except for knot A, the expansion times relative to the star from the transverse motion of the other two knots (Table 2) are consistent with the result for the arc center.

For the arcs there is thus evidence that the expansion time for the arc itself is less than the 
time since its ejection from the star. The possible causes of these different expansion times are discussed in \S {5}.

In summary, the knots, arcs etc. are {\it kinematically distinct} from the general expansion of the ejecta. 
The mean total space motion for the 10 knots in Table 3 is 113 $\pm$ 14  km s$^{-1}$ compared
to the 40 - 60 km s$^{-1}$ velocity of expansion inferred from the double-peaked profiles of the 
hydrogen and Ca II emission and the maser emission. Their mean space motion is due almost entirely to 
the mean transverse velocity of  112 $\pm$ 15 km s$^{-1}$ compared to only -5 $\pm$ 3 km s$^{-1}$ 
for the radial motion. Similarly,  the mean transverse velocity  for the remaining features in Table 2 is 89 $\pm$ 9 km s$^{-1}$.  In almost all cases with a radial velocity, the tranverse motions 
are significantly larger than the radial component. The motions of the discrete features are thus 
dominated by their transverse motions.

\section{Geometry of the Ejecta}

The morphology of IRC+10420's circumstellar ejecta has eluded previous studies. While the outer 
rings are consistent with  a basically spherical outflow from several thousand years ago perhaps when 
the star was in a different evolutionary state as a red supergiant, the inner ejecta is very 
complex and indicative of localized ejection events.  Suggestions for the orientation 
and geometry of the ejecta have ranged from an equatorial disk viewed 
at an angle to our line of sight (Paper I), a bipolar outflow (Oudmaijer et al 1994, Paper II,
Davies et al 2007) to an essentially spherical outflow (Paper III). 
Similarly, the various interpretations based on the strong OH maser emission include  a spherical outflow \citep{Bow84}, a bipolar outflow with a disk-like structure viewed edge-on (Diamond et al 1983), and a weakly bipolar, slightly oblate outflow with clumping \citep{Nedo92}.

Our results for the total space motions and resulting orientation for several of the  
discrete knots and condensations  (Table 3) interestingly show that while they have a range of
ejection velocities, and expansion ages they are all moving very close to the plane of the
sky or towards us by at most about 12$^{\circ}$ \footnote{The transverse velocity, expansion age, and orientation depend on the adopted distance of 5 kpc. A possible distance range 
of 4 -- 6 kpc corresponds to an uncertainty of up to20\% in the transverse velocity. The 
corresponding uncertainty in the time since ejection and the angle $\theta$ relative to the plane of the sky is small and significantly less than that due to the measurement errors.}.  With an optical thickness greater than one for IRC+10420's 
inner ejecta (Paper II), it is not  surprising that we do not find any features moving away from us. 
The similar orientation of most of these features, at different position angles and
ejected at different times over several hundred years, suggests that we are {\it viewing IRC+10420
nearly pole-on and therefore, looking nearly directly down onto its equatorial plane.}  
The SW fan, which  extends over an  arc covering approximately 70$^{\circ}$ projected onto the sky,   
 is a likely candidate to represent  the equatorial plane which would then be tilted by only 
 about 8$^{\circ}$ out of  the plane of the sky with the southwest side of the ejecta towards us.  
 The other features, the various knots and the 
 semi-circular arcs, would then all lie within $\approx$ 10$^{\circ}$ of the equatorial plane. 
Furthermore as noted above, the motions of the various knots etc are dominated by their
transverse velocities. They have little radial motion, supporting the interpretation that 
we are viewing these features essentially face-on.   

\citet{Davies07} have presented evidence in support of a bipolar outflow based on the velocites of
the reflected H$\alpha$ and Fe II emission across the nebula and argue for a preferred  axis of
symmetry at $\sim$ 45$^{\circ}$ on the basis of the optical and infrared images. The H$\alpha$ and Fe II
emission however show different kinematic patterns. The Fe II does not show a clear bipolar pattern; it has
radial gradient with lower velocities near the center and higher in the outer parts. H$\alpha$ shows 
the stongest evidence for bipolarity with lower velocities to the SW and higher velocities to the NE 
with a typical velocity difference of $\approx$ 10 km s$^{-1}$ within the inner region;  the strongest 
evidence for an outflow is beyond the inner 2$\arcsec$

We looked for any
evidence for an axis of symmetry in the motion of the knots especially between the SW and SE/NE quadrants
of the nebula\footnote{The semi-circular arcs were not included because of possible expansion.}.
For those few knots with radial motions, we find a small difference of 12 km s$^{-1}$ with the SW 
quadrant having a larger component of motion towards us. Given the apparent orientation of the ejecta,
 this is not surprising and may be due as much to the tilt of the SW side  towards our line of
 sight as to a bipolar outflow. Given the nearly pole-on geometry of the inner ejecta, the motions and 
 orientation of the various arcs and knots within about 2$\arcsec$  do not provide any direct information 
 on an axis of symmetry or bipolar outflow, and  the apparent velocity difference in the H$\alpha$ emission 
 with position in the inner nebula may also be due as much to its geometry as to an actual bipolar outflow.

 In our previous papers on the extreme red supergiant VY CMa \citep{Smi01,RMH05,RMH07,Smi04} we 
 have emphasized the presence of prominent arcs, knots and large looplike structures in its 
 very visible ejecta, all evidence for localized ejections which we have suggested are due 
 to large-scale surface activity and magnetic fields.  There are both similarities and important differences between 
 the circumstellar environments of VY CMa and IRC~+10420, although we are viewing them from different perspectives\footnote{VY CMa may be tilted $\approx$ 15$^{\circ}$--30$^{\circ}$ to the line
  of sight.}.  IRC~+10420's ejecta is apparently concentrated to the equatorial plane and it
  does not show the large prominence-like loops seen in VY CMa; however, the semi-circular arcs
 may be evidence for related structures.

\section{Discussion -- Mass Loss History }

The circumstellar ejecta of IRC~+10420 separates into the outer approximately spherical shells
 5$\arcsec$ -- 6$\arcsec$ from the star and the complex inner ejecta within 2$\arcsec$ of the 
 star. Adopting the nominal expansion velocity of $\approx$ 50 -- 60 km sec$^{-1}$ the 
 outer shells were ejected about 3000 years ago. Our transverse motion for the outer nebulosity
 about 4$\arcsec$ away confirms an expansion age of $\approx$ 2000 years. There is also evidence 
 for more distant associated 
 ejecta 8$\arcsec$ - 9$\arcsec$ away (Figure 5). This visual nebulosity  very likely corresponds to  
 the distant arc reported by \citep{KW95} from coronagraphic imaging in the near-infrared. 
 Assuming that this material was expelled with a 
 similar velocity, then it would have been ejected 4000--5000 years ago. These outer shells are similar to the shells or ejecta associated with many post-AGB stars and very likely result from pulsational 
 mass loss as a red supergiant in the case of IRC~+10420. In contrast, the inner 
 circumstellar material was apparently ejected at different times and in different directions
 beginning about 600 - 800 years ago up to fairly recently ending perhaps less than 100 years ago. 
 This result is consistent with other studies suggesting that a high mass loss period ended
 recently (Bl\"ocker et al. (1999), Paper III).
 Ejecta younger than 70-80 years however would be within the PSF dominated region and would not
 be included in our measurements. Thus we see at least two major epochs of mass loss separated by
 at least 1000 years or more. Furthermore, the results for the inner ejecta, suggest that the 
 more recent period was punctuated by times of increased activity. Many of the features have 
 expansion ages corresponding to 300 - 400 years ago and again from about 100 - 200 years ago.

%Arcs

As discussed in \S 3.8, the semi-circular arcs may be expanding  as well as traveling away 
from the star. An ejection time of 300 -- 400 years is consistent within the errors for all three
arcs, although it may be as high as 700 years for Arc 1. The results for Arcs 1 and 3 also indicate
expansion times much less than the time since ejection from the star. This could be due to the 
significant uncertainty in these measurements although the results for the knots in Arc 1 are
internally consistent.  As discussed in the next section, magnetic fields may play a role in 
in the origin of IRC~+10420's episodic mass loss.  They may also provide an explanation for the much
shorter measured expansion times for the arcs.  The strength of the magnetic field is 
estimated from the circular polarization of the OH masers on the inside edge of the maser shell
at about 7000 AU from the star \citep{Nedo92}, the average distance of the arcs (see Figure 6).  Suzuki (2007)  has demonstrated that  it is possible for hot bubbles ejected from red supergiants 
 to last much longer than their cool down times due to extended magnetic field lines from the 
 star which constrain the bubbles of gas as  they travel away from the star. While the models do
 not include stars as warm as IRC~+10420,  magnetic fields may restrict the expansion of a 
 bubble or loop. Given the apparent tilt of Arc 1, it also seems likely that the arcs are actually
 expanding loops  similar to the much larger structures in the ejecta of VY CMa.

\section{Concluding Remarks -- IRC~+10420, Convective Activity and Magnetic Fields}

The morphology of IRC~+10420's inner ejecta shows a recent  mass loss history dominated by localized 
ejection events  in random directions and at different times.  Mass loss and the winds of cool 
evolved stars including the AGB stars, red supergiant and red giants
are normally attributed to global  pulsation combined with radiation pressure on the dust which 
further drives the mass loss. These mechanisms can account for  relatively uniform essentially 
spherical ejecta like IRC~+10420's outer shells. But the complex and episodic mass loss evident in the 
inner regions require a non-uniform mass loss mechanism such as large-scale surface activity. 
There is now an increasing number of observations of  ``starspots'', 
large surface asymmetries, and outflows  associated with red supergiants, giants and AGB stars
\citep{Tut97,Mon04,Kiss09} consistent with a convective origin. Non-radial pulsations may be an alternative but are not consistent with the narrow looplike
structures observed, for example,  in the ejecta of VY CMa. Furthermore,  magnetic fields  
 associated with the maser emission are now confirmed in the ejecta of several of these stars including the strong OH/IR sources, VX Sgr, S Per, NML Cyg, and VY CMa \citep{Vlem02,Vlem04}.

Magnetic field strengths of from $\approx$ 1/6 to 15 mG have also been reported in the ejecta of IRC~+10420
from the observed circular polarization of the OH maser emission \citep{Reid79,Cohen87,Nedo92}. Figure 6
shows the distribution of the various maser sources superimposed on an image of IRC~+10420. The inner OH
emission at $\sim$ 1.5$\arcsec$ is the location of the circularly polarized emission and is coincident
with the inner ejecta. Adopting a conventional extrapolation ($B \; \propto \; r^{-2}$), 
a 1mG field at  r $\sim$ 5000 -- 7000 AU  would give B $\sim$ 3 kG  at the stellar surface, 
 high for a global field, and it would also exceed other local energy densities 
 (see Paper III). A field proportional 
to $r^{-1}$, however,  would give a surface magnetic energy density comparable to the thermal energy density
We also note that in the case of IRC~+10420, the knots and arcs appear to be concentrated in 
the equatorial region. We know that as a star evolves to warmer temperatures the increased dynamical instability will be 
enhanced in the equatorial region as the star's rotation also increases. Thus for IRC~+10420 we may be 
observing the combined effects of turbulence/convection and increased rotation. 

The case for the role of convection and magnetic fields on the mass lass history and mechanism in
evolved massive stars is of course strongest in the red supergiant stage. IRC~+10420 is a post-red
supergiant and the inner ejecta were apparently formed long after it had been a red supergiant.
However, it is also close to the upper luminosity boundary for evolved stars in the H-R diagram and to the
critical temperature regime, 7000 -- 9000$^{\circ}$K (deJager 1998), where dynamical instabilities
 become important for stars evolving to warmer temperatures.

In summary, the evidence for episodic mass loss events associated
with convective/magnetic activity is visible in the resolved circumstellar ejecta of objects 
like VY CMa and IRC~+10420 with their numerous knots and arcs ejected at different times from 
separate regions on the star's surface. The role of these dynamical instabilities on the mass loss
histories of the most luminous evolved stars must be considered as a probable source of 
their high mass loss episodes especially in their final stages.

\acknowledgments
The authors  thank L. Andrew Helton for advice on data reduction.  This work was supported by NASA
through grant GO-11180 from the Space Telescope Science Institute.


%% in parentheses, after the keyword, but they will not be verified.

{\it Facilities:}  \facility{HST (WFPC2)}.



\appendix

\section{Angular Measurements} 

The measurements for the individual features in the different filter combinations with the 
number of independent measurements are summarized in Table A1. 


\clearpage

%Tables
\clearpage

%Table 1
%\documentclass{aastex}
%\begin{document}

\begin{deluxetable}{lllll}
\tablecolumns{5}
\tablecaption{List of Observations and Combined Images}
\tablehead{
\colhead{Date}  &  \colhead{Filter}  &  \colhead{Exposure Times (seconds)} &  \colhead{Combined Images}}
\startdata
April 5, 1996 & F467M   &   12${s}$, 30${s}$ $\times$ 2, 140${s}$ $\times$ 2  & F467{\it s}1 (72${s}$),  F467{\it l}1 (280${s}$)\\
(Epoch 1) &  F547M & .5${s}$, 3${s}$, 10${s}$, 40${s}$, {\it 140}${s}$ $\times$ 2 \tablenotemark{a} & F547{\it s}1 (13.5${s}$), F547{\it l}1 (320${s}$) \tablenotemark{b}\\
March 9, 2008  &  F467M   &   12${s}$, 30${s}$ $\times$ 2, 140${s}$ $\times$ 2 & F467{\it s}2 (72${s}$),  F467{\it l}2 (280${s}$)\\
(Epoch 2) &  F547M & .5${s}$, 3${s}$, 10${s}$ $\times$ 2, 40${s}$, 300${s}$  & F547{\it s}2 (23.5${s}$),  F547{\it l}2 (340${s}$) \tablenotemark{c} \\
  &  F675W   &   .5${s}$, 5${s}$, 30${s}$ $\times$ 3, 600${s}$ & F675{\it s} (5.5${s}$),  F675{\it l} (90${s}$) \\
  &  &  & F675{\it ex} (600${s}$)\\
  &  F1042M  &   .5${s}$, 5${s}$, 30${s}$, 100${s}$ & F1042{\it s} (5.5${s}$),  F1042{\it l} (130${s}$)\\

\enddata
\tablenotetext{a}{The 140 second exposures here were not dithered; see \S{2.1}}
\tablenotetext{b}{Filter 547 didn't have matching times between the two epochs, so the short and long combined exposures are not the same times between epochs.}
\tablenotetext{c}{The 300s image was combined with the short 40s image to help remove cosmic rays.  The images from Epoch 1 were combined to match the total exposure time as closely as possible for both the short and long exposures.}
\tablecomments{All of the images where two or more exposures were taken in the same epoch with the same time where taken using the HST dither technique (except as noted above).  The images were offset by 2.5 pixels (short exposures) or 5 pixels (long exposures).}
\end{deluxetable}

%\end{document}
%\endinput
\clearpage

%Table 2
%\documentclass{aastex}
%\begin{document}
\begin{deluxetable}{lccccc}
\tabletypesize{\scriptsize}
\tablecaption{The Position, Transverse Velocity and Direction of Motion Relative to the Star} 
\tablewidth{0pt}
\tablehead{
 & \colhead{Radial Distance} & \colhead{Position Angle} &  \colhead{V$_{\it Trans}$}  & \colhead{$\phi$} & \colhead{Expansion Age} \\
\colhead{Feature ID} & \colhead{(arcsec)} & \colhead{(deg)}  &  \colhead{(km s$^{-1}$)} & \colhead{(deg)} & \colhead{(yr)} \\
}
\startdata

SE Jet(1), Knot A\tablenotemark{a} & 0.57 & 122 $\pm$ 3 & 17 $\pm$ 10  & 127 $\pm$ 10 & 800$^{+1100}_{-300}$\\
SE Jet(1), Knot B & 0.69 & 126 $\pm$ 3 & 27 $\pm$ 9  & 140 $\pm$ 12 & 600$^{+300}_{-150}$\\
SE Jet(2), Knot A\tablenotemark{a} & 0.38 & 175 $\pm$ 3 & 126 $\pm$ 6  & 144 $\pm$ 3 & 83$^{+8}_{-7}$\\
SE Knot A & 0.57 & 136 $\pm$ 3 & 89 $\pm$ 7  & 180 $\pm$ 3 & 160$^{+15}_{-10}$\\
SE Knot B & 1.02 & 152 $\pm$ 2 & 42 $\pm$ 14  & 80 $\pm$ 16 & 600$^{+310}_{-150}$\\
SE Knot C & 1.17 & 137 $\pm$ 3 & 118 $\pm$ 20  & 11 $\pm$ 3 & 230$^{+50}_{-35}$\\
SE Knot D & 1.35 & 138 $\pm$ 3 & 60 $\pm$ 12  & 104 $\pm$ 8 & 550$^{+130}_{-90}$\\
SE Knot E & 1.58 & 141 $\pm$ 2 & 54 $\pm$ 5  & -139 $\pm$ 2 & 700$^{+70}_{-60}$\\
SE Knot F & 1.25 & 149 $\pm$ 2 & 69 $\pm$ 13  & 137 $\pm$ 6 & 450$^{+110}_{-70}$\\
SE Knot G & 1.39 & 148 $\pm$ 2 & 98 $\pm$ 7  & 161 $\pm$ 2 & 350$^{+30}_{-25}$\\
\vspace{4 mm}
SE Outer Knot & 4.01 & 143 $\pm$ 2 & 49 $\pm$ 10 & 122 $\pm$ 5 & 2000$^{+500}_{-230}$\\

SW Jet, Knot B\tablenotemark{a} & 0.54 & -173 $\pm$ 2 & 218 $\pm$ 9 & -165 $\pm$ 3 & 70$^{+5}_{-5}$\\
SW Square, Knot A & 1.06 & -125 $\pm$ 2 & 131 $\pm$ 5 & 133 $\pm$ 3 & 190$^{+10}_{-10}$\\
SW Square, Knot B & 1.07 & -137 $\pm$ 3 & 128 $\pm$ 10 & -167 $\pm$ 3 & 200$^{+20}_{-15}$\\
SW Square, Knot C\tablenotemark{a} & 1.24 & -136 $\pm$ 2 & 96 $\pm$ 21 & -160 $\pm$ 8 & 320$^{+80}_{-60}$\\
SW Square, Knot D & 1.30 & -132 $\pm$ 2 & 89 $\pm$ 15 & -174 $\pm$ 6 & 360$^{+70}_{-50}$\\
SW Knot A & 1.36 & -146 $\pm$ 3 & 41 $\pm$ 6 & -80 $\pm$ 4 & 800$^{+150}_{-100}$\\
SW Triangle, Knot A & 1.18 & -159 $\pm$ 2 & 214 $\pm$ 22 & -145 $\pm$ 6 & 140$^{+20}_{-15}$\\
SW Triangle, Knot B & 1.26 & -165 $\pm$ 2 & 15 $\pm$ 9 & -6 $\pm$ 18 & \nodata \tablenotemark{b}\\
SW Triangle, Knot C & 1.46 & -155 $\pm$ 3 & 50 $\pm$ 8 & -109 $\pm$ 8 & 700$^{+125}_{-100}$\\
SW Fan, Knot A & 1.99 & -180 $\pm$ 3 & 104 $\pm$ 31 & -174 $\pm$ 16 & 470$^{+200}_{-100}$\\
\vspace{4 mm}
SW Fan, Knot B\tablenotemark{a} & 1.89 & -135 $\pm$ 2 & 103 $\pm$ 25 & -149 $\pm$ 10 & 450$^{+140}_{-90}$\\

E Jet, Knot B & 0.66  & 85 $\pm$ 2 & 165 $\pm$ 17 & 61 $\pm$ 5 & 95$^{+15}_{-10}$\\
E Jet, Knot C & 0.82  & 89 $\pm$ 2 & 136 $\pm$ 5 & 161 $\pm$ 4 & 150$^{+10}_{-10}$\\
\vspace{4 mm}
NE Knot G & 0.85 & 16 $\pm$ 2 & 94 $\pm$ 16 & -148 $\pm$ 6 & 200$^{+50}_{-35}$\\

Arc 1, Knot A & 1.32 & 67 $\pm$ 2 & 50 $\pm$ 7 & -180 $\pm$ 6 & 650$^{+120}_{-90}$\\
Arc 1, Knot B\tablenotemark{a} & 1.33 & 57 $\pm$ 2 & 116 $\pm$ 5 & -173 $\pm$ 3 & 270$^{+10}_{-10}$\\
Arc 1, Knot C\tablenotemark{a} & 1.64 & 57 $\pm$ 2 & 108 $\pm$ 20 & 81 $\pm$ 4 & 380$^{+80}_{-60}$\\
Arc 1, Knot D & 1.71 & 65 $\pm$ 2 & 130 $\pm$ 5 & 158 $\pm$ 2 & 300$^{+50}_{-40}$\\
Arc 1, Knot E & 1.68 & 67 $\pm$ 2 & 69 $\pm$ 14 & 139 $\pm$ 7 & 570$^{+150}_{-100}$\\
\vspace{4 mm}
Arc 1, Knot G\tablenotemark{a} & 1.50 & 73 $\pm$ 2 & 107 $\pm$ 13 & 116 $\pm$ 3 & 330$^{+50}_{-40}$\\

Arc 2, Knot A & 1.50 & 78 $\pm$ 2 & 110 $\pm$ 25 & 166 $\pm$ 11 & 330$^{+100}_{-80}$\\
Arc 2, Knot B & 1.84 & 81 $\pm$ 2 & 62 $\pm$ 21 & -155 $\pm$ 13 & 700$^{+400}_{-200}$\\
Arc 2, Knot C & 1.98 & 84 $\pm$ 2 & 87 $\pm$ 15 & 141 $\pm$ 6 & 550$^{+100}_{-80}$\\
Arc 2, Knot D & 1.87 & 89 $\pm$ 2 & 88 $\pm$ 10 & 150 $\pm$ 6 & 500$^{+80}_{-60}$\\
Arc 2, Knot E & 1.59 & 86 $\pm$ 2 & 119 $\pm$ 25 & 134 $\pm$ 8 & 300$^{+100}_{-70}$\\
\vspace{4 mm}
Arc 2, Knot F\tablenotemark{a} & 1.64 & 73 $\pm$ 2 & 97 $\pm$ 13 & 148 $\pm$ 9 & 400$^{+90}_{-70}$\\

Arc 3, Knot A\tablenotemark{c} & 1.66 & 104 $\pm$ 2 & 23 $\pm$ 17 & -86 $\pm$ 3 & 1700$^{+1000}_{-500}$\\
Arc 3, Knot B & 1.77 & 114 $\pm$ 2 & 102 $\pm$ 13 & 158 $\pm$ 6 & 420$^{+40}_{-30}$\\
Arc 3, Knot C & 1.51 & 117 $\pm$ 2 & 199 $\pm$ 14 & 111 $\pm$ 3 & 200$^{+20}_{-15}$\\

\enddata
\tablenotetext{a}{These features also have radial velocities and are listed in Table 3.}
\tablenotetext{b}{The transverse velocity is too small to find a meaningful expansion age.}
\tablenotetext{c}{This knot is very close to a diffraction spike which causes the high error the velocity and age.}
\end{deluxetable}

%\end{document}
%\endinput
\clearpage

%Table 3
%\documentclass{aastex}
%\begin{document}
\begin{deluxetable}{lccccc}
\tablecaption{Summary of Vector Motions and Ejection Ages for the Major Features}
\tablewidth{0pt}
\tablehead{
\colhead{Feature} & \colhead{V$_{\it Trans}$}\tablenotemark{a} & \colhead{V$_{\it R}$}  & \colhead{V$_{\it Total}$} & \colhead{$\theta$} & \colhead{Expansion Age}\\
& \colhead{(km s$^{-1}$)}  &  \colhead{(km s$^{-1}$)}  &  \colhead{(km s$^{-1}$)}  &  \colhead{(deg)} & \colhead{(yr)} 
}
\startdata
East Jet, Knot C  & 136  & -5.9  & 136 $\pm$ 10 & -2.5 $\pm$ 2 & 150$^{+12}_{-10}$ \\
SE Jet (1), Knot A  & 18 & -3.6 & 17 $\pm$ 12 & -12 $\pm$ 10.6 & 800$^{+1000}_{-300}$\\
SE Jet (2), Knot A  & 126  & -3.8 &  126 $\pm$ 11 & -1.7 $\pm$ 1.4 & 80$^{+8}_{-7}$\\
SW Jet, Knot B  & 218  & -5.6 & 218 $\pm$ 12 & -1.5 $\pm$ 1.3 & 70$^{+8}_{-7}$\\
SW Sq, Knot C  & 96  & -19.5 & 98 $\pm$ 21 & -11.5 $\pm$ 4 & 320$^{+90}_{-60}$\\
SW Fan, Knot B  & 103  & -15.3 & 105 $\pm$ 25 & -8.4 $\pm$ 3.8 & 450$^{+140}_{-90}$\\
Arc 1, Knot B & 116  & 16.3 & 117 $\pm$ 16 & 8 $\pm$ 3 & 270$^{+40}_{-30}$\\
Arc 1, Knot C & 108  & 3.3 & 108 $\pm$ 19 & 1.8 $\pm$ 1.2 & 380$^{+80}_{-60}$\\
Arc 1, Knot G & 107 & -6.5 & 107 $\pm$ 13 & -3.5 $\pm$ 2.5 & 330$^{+50}_{-40}$\\
Arc 2, Knot F & 97 &  -9 & 98 $\pm$ 13 & -5.3 $\pm$ 3.2 & 400$^{+50}_{-40}$\\

\enddata 
\tablenotetext{a}{From Table 2.}
\tablecomments{For transverse velocity errors see Table 2.  Errors in V$_{R}$ are assumed to be on the order of 3 km s$^{-1}$ for calculation purposes.  They are much less than the transverse velocity errors.}

\end{deluxetable}

%\end{document}
%\endinput

\clearpage

%Table 7
%\documentclass{aastex}
%\begin{document}
\begin{deluxetable}{lccccccc}
\tablecaption{Arc Center Point Fit Information}
\tablewidth{0pt}
\tablehead{
& \colhead{Position Angle} & \colhead{V$_{\it Trans}$} & \colhead{$\phi$} & \colhead{Expansion} & \colhead{V$_{\it R}$} & \colhead{V$_{\it Total}$} & \colhead{$\theta$} \\
\colhead{Arc} & \colhead{(deg)} & \colhead{(km s$^{-1}$)} & \colhead{(deg)} & \colhead{Age (yr)}\tablenotemark{a} & \colhead{(km s$^{-1}$)} & \colhead{(km s$^{-1}$)} & \colhead{(deg)} \\
}
\startdata
Arc 1 & 64 $\pm$ 5 & 51 $\pm$ 30\tablenotemark{b} & 153 $\pm$ 10 & 700$^{+1000}_{-400}$ & -0.4 & 51 $\pm$ 30 & -0.4 $\pm$ 0.8 \\
Arc 2 & 84 $\pm$ 5 & 107 $\pm$ 40\tablenotemark{b} & 155 $\pm$ 8 & 370$^{+200}_{-100}$ & \nodata & \nodata & \nodata \\
Arc 3 & 111 $\pm$ 5 & 140 $\pm$ 50\tablenotemark{b} & 121 $\pm$ 10 & 300$^{+200}_{-100}$ & \nodata & \nodata & \nodata \\

\enddata

\tablenotetext{a}{This expansion age is for the center of the arc relative to the star, and not for the individual knots.}
\tablenotetext{b}{The errors are from the spread in velocities found with the different fits to each arc.}
\tablecomments{Values listed are for the arcs are for the best fits only.}

\end{deluxetable}

%\end{document}
%\endinput

\clearpage

%Table 3
%\documentclass{aastex}
%\begin{document}
\begin{deluxetable}{lccccc}
\tabletypesize{\scriptsize}
\tablecaption{The Position, Transverse Velocity and Direction of Motion Relative to the Arc Center} 
\tablewidth{0pt}
\tablehead{
 & \colhead{Radial Distance} & \colhead{Position Angle} & \colhead{V$_{\it Trans}$ } & \colhead{$\phi$} & \colhead{Expansion Age} \\
\colhead{Feature ID} & \colhead{(arcsec)} & \colhead{(deg)}  &  \colhead{(km s$^{-1}$)} & \colhead{(deg)} & \colhead{(yr)}\\
}
\startdata
Arc 1, Knot A & 0.19 & -132 $\pm$ 2 & 23 $\pm$ 15 & -97 $\pm$ 4 & 200$^{+900}_{-100}$\\
Arc 1, Knot B & 0.23 & -66 $\pm$ 2 & 79 $\pm$ 15 & -173 $\pm$ 2 & 70$^{+100}_{-50}$\\
Arc 1, Knot C & 0.27 & -4 $\pm$ 2 & 99 $\pm$ 15 & 51 $\pm$ 3 & 65$^{+100}_{-50}$\\
Arc 1, Knot D & 0.22 & 69 $\pm$ 2 & 44 $\pm$ 15 & -20 $\pm$ 3 & 120$^{+400}_{-70}$\\
Arc 1, Knot E & 0.19 & 88 $\pm$ 2 & 24 $\pm$ 18 & 107 $\pm$ 6 & 190$^{+900}_{-100}$\\
\vspace{4 mm}
Arc 1, Knot G & 0.20 & 169 $\pm$ 3 & 74 $\pm$ 15 & 90 $\pm$ 3 & 65$^{+100}_{-50}$\\
Arc 2, Knot A & 0.25 & -50 $\pm$ 3 & 21 $\pm$ 13 & -62 $\pm$ 9 & 280$^{+300}_{-100}$\\
Arc 2, Knot B & 0.24 & 57 $\pm$ 3 & 83 $\pm$ 21 & -60 $\pm$ 2 & 70$^{+150}_{-40}$\\
Arc 2, Knot C & 0.31 & 84 $\pm$ 3 & 31 $\pm$ 15 & 18 $\pm$ 4 & 235$^{+1000}_{-100}$\\
Arc 2, Knot D & 0.21 & 122 $\pm$ 3 & 20 $\pm$ 12 & -5 $\pm$ 26 & 250$^{+1000}_{-100}$\\
Arc 2, Knot E & 0.25 & -165 $\pm$ 3 & 42 $\pm$ 15 & -71 $\pm$ 13 & 150$^{+1000}_{-100}$\\
\vspace{2 mm}
Arc 2, Knot F & 0.36 & -17 $\pm$ 15 & 21 $\pm$ 12 & -27 $\pm$ 15 & 400$^{+400}_{-100}$\\
Arc 3, Knot A & 0.24 & 33 $\pm$ 3 & 160 $\pm$ 15 & -63 $\pm$ 5 & 36$^{+50}_{-20}$\\
Arc 3, Knot B & 0.13 & 144 $\pm$ 3 & 85 $\pm$ 15 & 106 $\pm$ 6 & 37$^{+60}_{-30}$\\
Arc 3, Knot C & 0.21 & -130 $\pm$ 3 & 68 $\pm$ 15 & 90 $\pm$ 5 & 74$^{+200}_{-40}$\\

\enddata

\end{deluxetable}

%\end{document}
%\endinput

\clearpage

%Table 5
%\documentclass{aastex}
%\begin{document}
\begin{deluxetable}{lcccccc}
\tablecaption{The 3D Motion of Arc 1 With Respect to the Arc Center}
\tablewidth{0pt}
\tablehead{
& \colhead{Position Angle} & \colhead{V$_{\it Trans}$} & \colhead{$\phi$} & \colhead{V$_{\it R}$\tablenotemark{a}} & \colhead{$\theta$} & \colhead{V$_{\it Total}$\tablenotemark{b}} \\
\colhead{Feature ID} & \colhead{(deg)} & \colhead{(km s$^{-1}$)} & \colhead{(deg)} & \colhead{(km s$^{-1}$)} & \colhead{(deg)} & \colhead{(km s$^{-1}$)} \\
}
\startdata
Arc 1, Knot A & -132 $\pm$ 2 & 23 $\pm$ 10 & -97 $\pm$ 7 & \nodata & \nodata & $\geq$ 23\\
Arc 1, Knot B & -66 $\pm$ 2 & 79 $\pm$ 11 & -173 $\pm$ 7 & 16.7 & 12 $\pm$ 3 & 81 $\pm$ 13\\
Arc 1, Knot C & -4 $\pm$ 2 & 99 $\pm$ 13 & 51 $\pm$ 7 & 3.7 & 2.1 $\pm$ 3 & 99 $\pm$ 14\\
Arc 1, Knot D & 69 $\pm$ 3 & 44 $\pm$ 12 & -20$\pm$ 7 & \nodata & \nodata & $\geq$ 44\\
Arc 1, Knot E & 88 $\pm$ 2 & 24 $\pm$ 11 & 107 $\pm$ 7 & \nodata & \nodata & $\geq$ 24\\
Arc 1, Knot G & 169 $\pm$ 3 & 74 $\pm$ 12 & 90 $\pm$ 7 & -6.1 & -4.8 $\pm$ 3 & 74 $\pm$ 14\\

\enddata
\tablenotetext{a}{These are the radial velocities relative to the value at the arc center (see \S3.8).  Negative values indicate the knot was moving faster than the arc center and coming towards us with respect to the arc center.}
\tablenotetext{b}{These errors are based soley on the measurement values and don't include errors from the measurement of the arc center.}

\end{deluxetable}

%\end{document}
%\endinput

\clearpage

%Table A1
%\documentclass{aastex}
%\begin{document}

\begin{deluxetable}{lcccc} 
\tabletypesize{\scriptsize}
\tablenum{A1}
\tablecaption{Angular Shifts in Arc Seconds for the 4 Filter Combinations} 
\tablewidth{0pt}
\tablehead{
\colhead{Feature ID} & \colhead{F467{\it s}} & \colhead{F467{\it l}} &  \colhead{F547{\it s}}  & \colhead{F547{\it l}}
} 
\startdata
SE Jet(1), Knot A & \nodata & \nodata & \nodata & 0.009 $\pm$ 0.003 (2) \\
SE Jet(1), Knot B & \nodata & \nodata & \nodata & 0.014 $\pm$ 0.005 (3) \\
SE Jet(2), Knot A & \nodata & \nodata & 0.051 $\pm$ 0.018 (2) & 0.063 $\pm$ 0.005 (3) \\
SE Knot A & \nodata & \nodata & \nodata & 0.045 $\pm$ 0.003 (3) \\
SE Knot B & \nodata & \nodata & \nodata & 0.021 $\pm$ 0.007 (3) \\
SE Knot C & \nodata & \nodata & \nodata & 0.059 $\pm$ 0.010 (3) \\
SE Knot D & \nodata & \nodata & \nodata & 0.030 $\pm$ 0.006 (3) \\
SE Knot E & \nodata & \nodata & \nodata & 0.027 $\pm$ 0.003 (3) \\
SE Knot F & \nodata & \nodata & \nodata & 0.034 $\pm$ 0.007 (3) \\
SE Knot G & \nodata & \nodata & \nodata & 0.049 $\pm$ 0.003 (2) \\
\vspace{4 mm}
SE Outer Knot & \nodata & \nodata & \nodata & 0.024 $\pm$ 0.003 (2) \\
SW Jet, Knot B & 0.101 $\pm$ 0.013 (3) & 0.103 $\pm$ 0.004 (3) & \nodata & 0.109 $\pm$ 0.005 (2) \\
SW Square, Knot A & \nodata & \nodata & \nodata & 0.066 $\pm$ 0.003 (3) \\
SW Square, Knot B & \nodata & \nodata & \nodata & 0.064 $\pm$ 0.004 (3) \\
SW Square, Knot C & \nodata & \nodata & 0.051 $\pm$ 0.014 (2) & 0.046 $\pm$ 0.007 (3) \\
SW Square, Knot D & \nodata & \nodata & \nodata & 0.045 $\pm$ 0.006 (3) \\
SW Knot A & \nodata & \nodata & \nodata & 0.021 $\pm$ 0.003 (3) \\
SW Triangle, Knot A & \nodata & \nodata & 0.106 $\pm$ 0.010 (2) & 0.110 $\pm$ 0.005 (2) \\
SW Triangle, Knot B & \nodata & \nodata & \nodata & 0.008 $\pm$ 0.004 (3) \\
SW Triangle, Knot C & \nodata & \nodata & \nodata & 0.025 $\pm$ 0.004 (3) \\
SW Fan, Knot A & \nodata & \nodata & 0.054 $\pm$ 0.013 (2) & 0.051 $\pm$ 0.011 (2) \\
\vspace{4 mm}
SW Fan, Knot B & \nodata & \nodata & \nodata & 0.052 $\pm$ 0.012 (3) \\
E Jet, Knot B & 0.083 $\pm$ 0.008 (3) & \nodata & \nodata & \nodata \\
E Jet, Knot C & \nodata & \nodata & \nodata & 0.068 $\pm$ 0.004 (3) \\
\vspace{4 mm}
NE Knot G & \nodata & 0.041 $\pm$ 0.008 (2) & \nodata & 0.049 $\pm$ 0.006 (3) \\
Arc 1, Knot A & \nodata & \nodata & \nodata & 0.024 $\pm$ 0.003  (3)\\
Arc 1, Knot B & \nodata & \nodata & \nodata & 0.058 $\pm$ 0.003  (3)\\
Arc 1, Knot C & \nodata & \nodata & \nodata & 0.051 $\pm$ 0.009  (3)\\
Arc 1, Knot D & \nodata & \nodata & \nodata & 0.070 $\pm$ 0.004  (3)\\
Arc 1, Knot E & \nodata & \nodata & \nodata & 0.035 $\pm$ 0.007  (3)\\
\vspace{4 mm}
Arc 1, Knot G & \nodata & \nodata & \nodata & 0.054 $\pm$ 0.006  (3)\\
Arc 2, Knot A & \nodata & \nodata & \nodata & 0.055 $\pm$ 0.009  (3)\\
Arc 2, Knot B & \nodata & \nodata & \nodata & 0.031 $\pm$ 0.011  (3)\\
Arc 2, Knot C & \nodata & \nodata & \nodata & 0.043 $\pm$ 0.007  (3)\\
Arc 2, Knot D & \nodata & \nodata & \nodata & 0.044 $\pm$ 0.005  (3)\\
Arc 2, Knot E & \nodata & \nodata & \nodata & 0.060 $\pm$ 0.010  (3)\\
\vspace{4 mm}
Arc 2, Knot F & \nodata & \nodata & \nodata & 0.049 $\pm$ 0.006  (3)\\
Arc 3, Knot A & \nodata & \nodata & \nodata & 0.011 $\pm$ 0.005  (3)\\
Arc 3, Knot B & \nodata & \nodata & \nodata & 0.051 $\pm$ 0.009  (3)\\
Arc 3, Knot C & \nodata & \nodata & \nodata & 0.100 $\pm$ 0.007  (3)\\

\enddata
\tablecomments{Numbers in parenthesis indicate the number of measurements that were used to find the angular shift in each filter combination.}
\end{deluxetable}
%\end{document}

%\endinput
\clearpage
%Figures

%figure 1 
\begin{figure}
\plotone{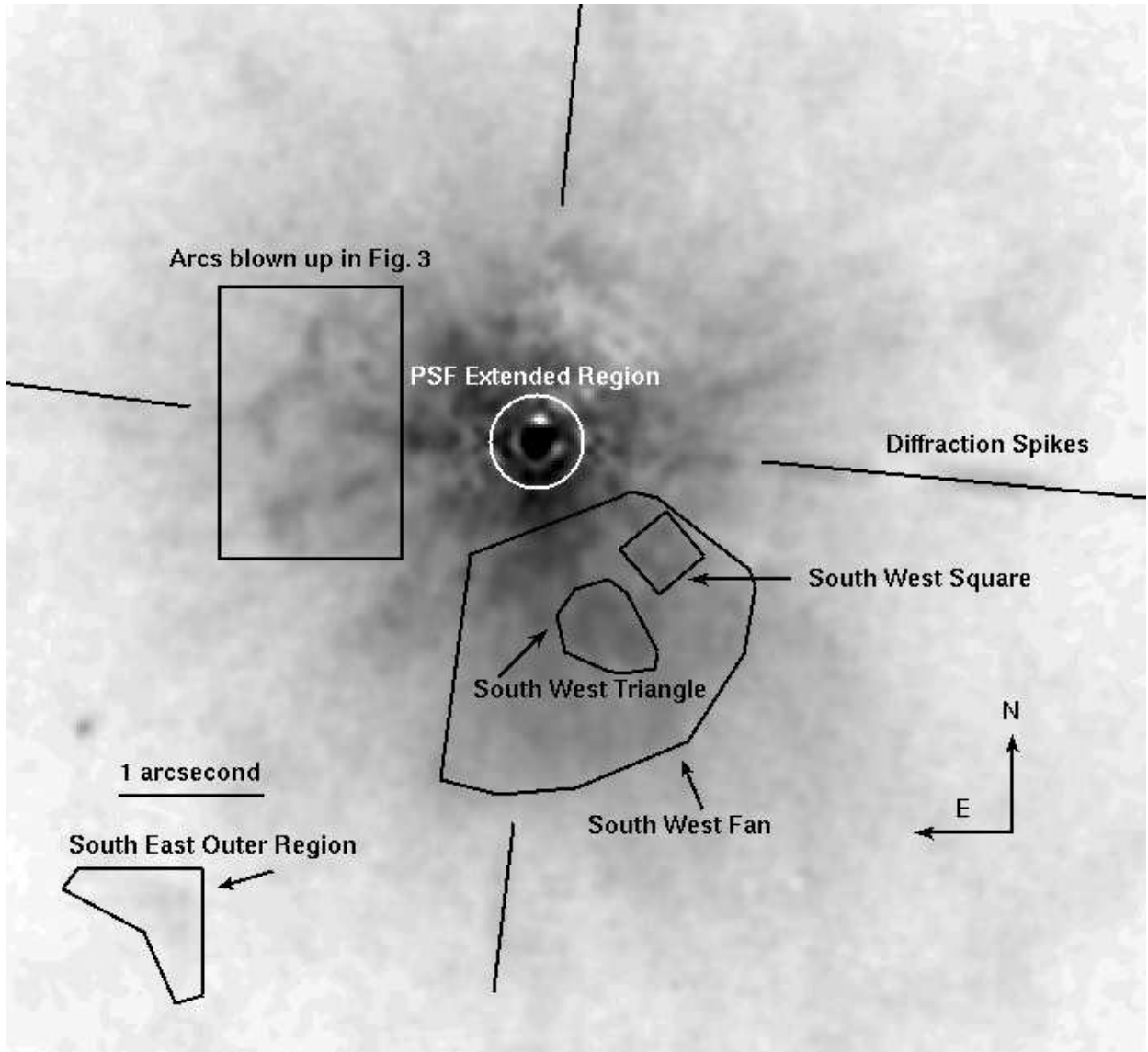}
\caption{Image 547s1 (see Table 1) with the regions of the image labeled.  1 arcsecond is roughly 10,000 AU in this image.  See Figure 2a and 2b for detailed labeling.}
\end{figure}

\clearpage

%figure 2 
\begin{figure}
\plotone{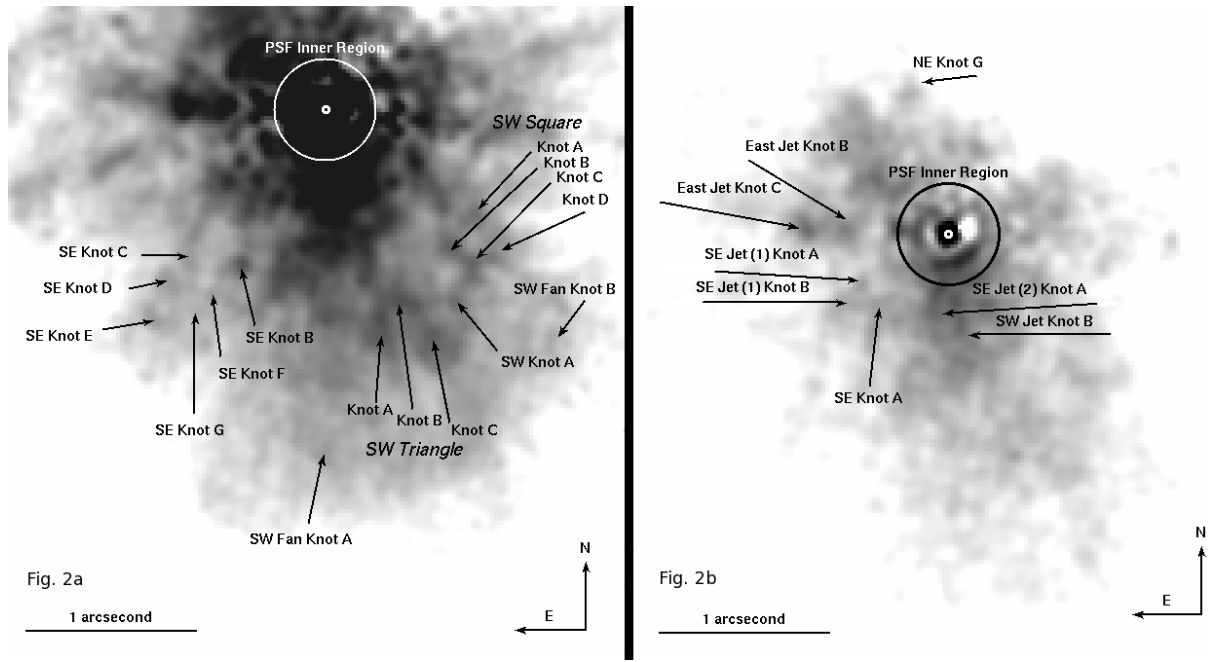}
\caption{Figure 2a (right) shows knots that are farther from the star center (Image 547s1).  Figure 2b shows the knots closer to the star (Image 467l1).}
\end{figure}

\clearpage 

%figure 3
\begin{figure}
\epsscale{0.75}
\plotone{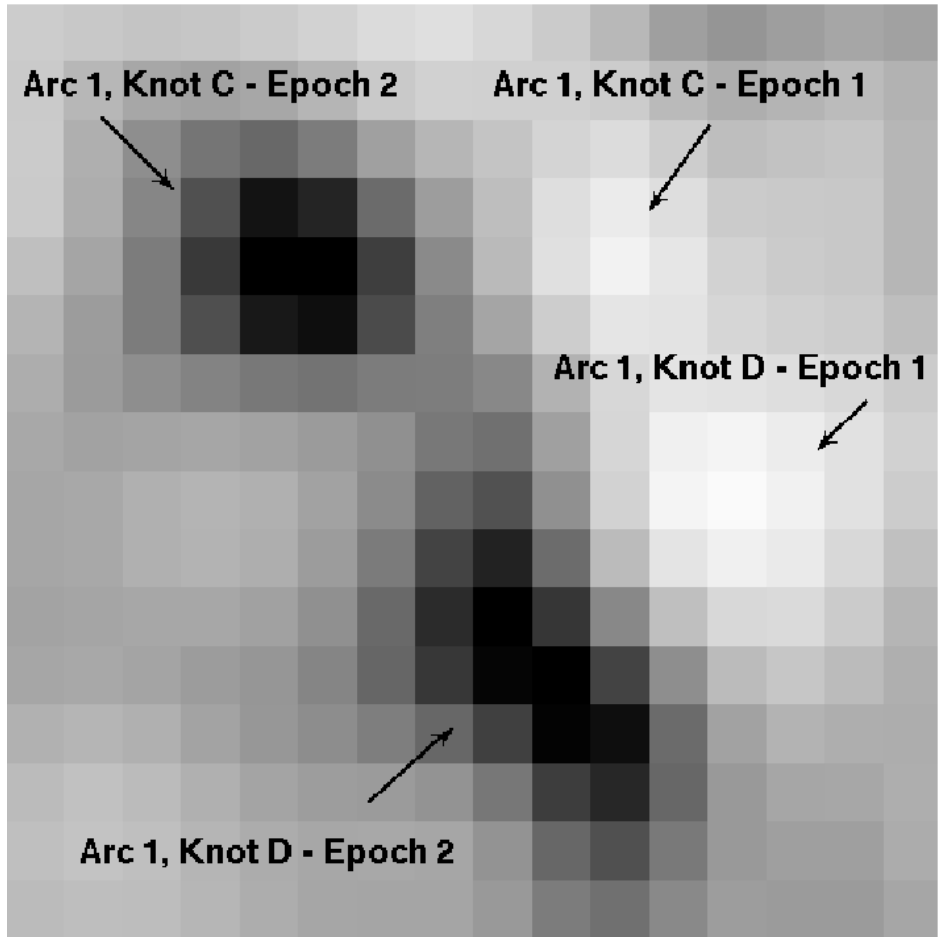}
\caption{An enlargement of a portion of Arc 1 to illustrate the motion of Knots C and D between the two epochs. The epoch 2 image is subtracted from epoch 1. This area does not have a complex background, so the offset shows relatively clearly using subtraction.  
The box is 0.36$\arcsec$ across.}
\end{figure}

\clearpage

%figure 4 
\begin{figure}
\epsscale{0.5}
\plotone{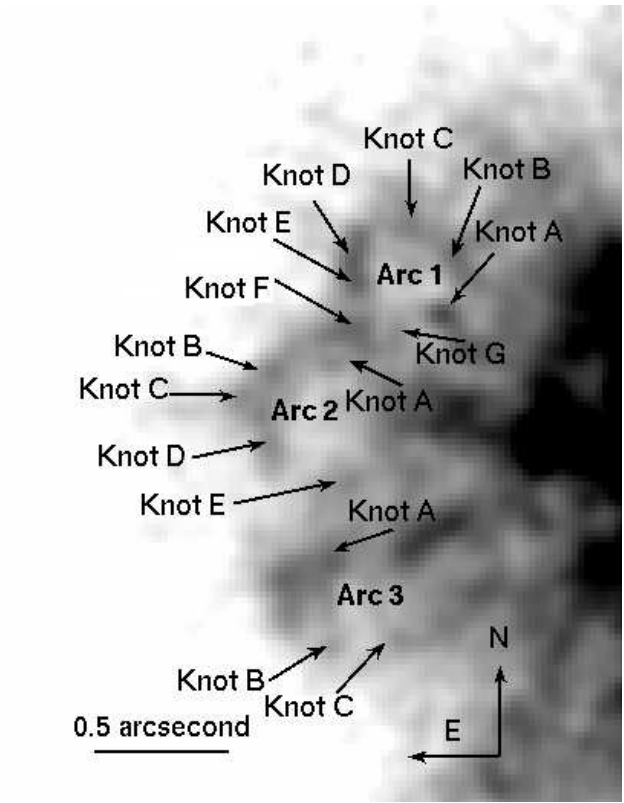}
\caption{Image 547l1 with the knots of the 3 arcs labeled (blow up of region shown in Figure 1).}
\end{figure}

\clearpage

%figure 5 
\begin{figure}
\epsscale{1.3}
\plotone{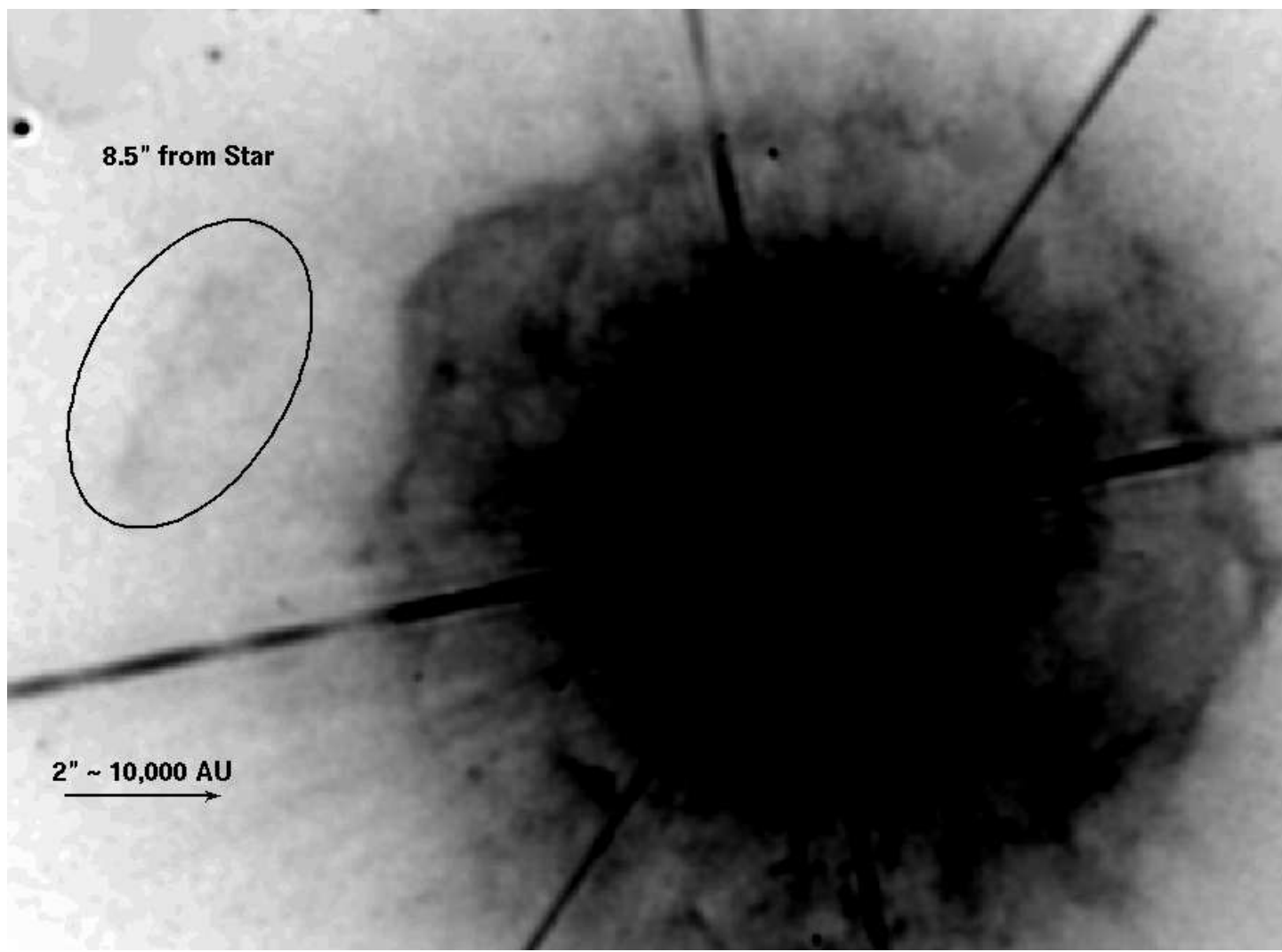}
\caption{The long exposure red image showing the outermost nebulosity at  8$\arcsec$ to  9$\arcsec$ from 
the central star.}
\end{figure}

\clearpage

%figure 6
\begin{figure}
\epsscale{1.0}
\plotone{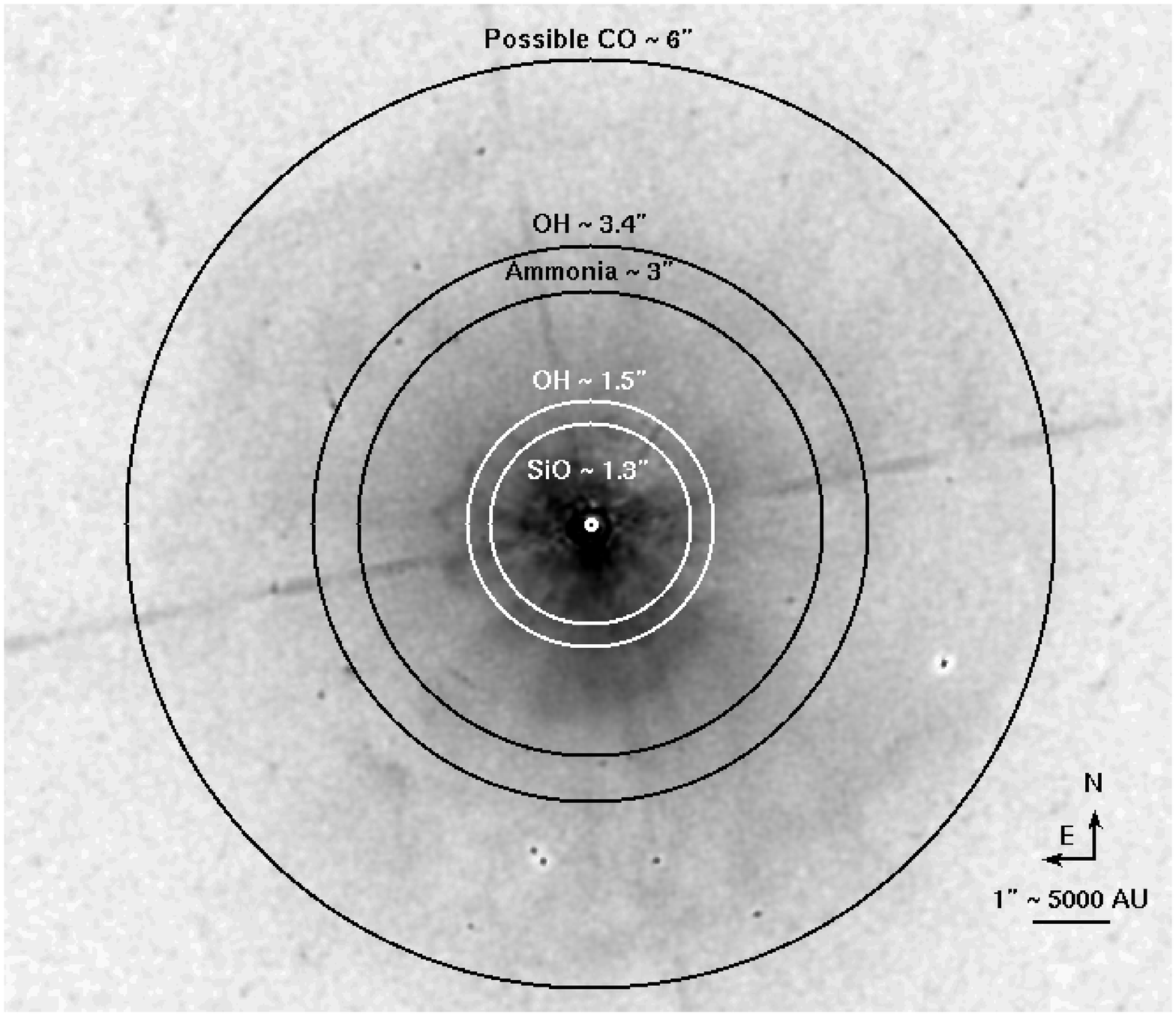}
\caption{Image 547l1 with the peak maser emissions overlayed:  SiO \citep{Castro01}, the 
inner OH ring \citep{Nedo92}, ammonia \citep{Ment95}, and the outer CO emission \citep{Castro07}} 
\end{figure}

%% The following command ends your manuscript. LaTeX will ignore any text

\end{document}